\documentclass[twoside,12pt]{article}
\usepackage{graphicx}
\begin{document}
\centerline{\bf Phase transition in hierarchy model of Bonabeau et al}
\centerline{\bf }

\bigskip
Dietrich Stauffer$^1$

\bigskip
Laboratoire PMMH, Ecole Sup\'erieure de Physique et Chimie Industrielle, 10
rue Vauquelin, F-75231 Paris, Euroland

\medskip
\noindent
$^1$ Visiting from Institute for Theoretical Physics, Cologne University, 
D-50923 K\"oln, Euroland; stauffer@thp.uni-koeln.de

\bigskip
The model of Bonabeau explains the emergence of social hierarchies
from the memory of fights in an initially egalitarian
society. Introducing a feedback from the social inequality into the
probability to win a fight, we find a sharp transition between
egalitarian society at low population density and hierarchical society
at high population density.

\medskip



\bigskip
How do social hierarchies emerge in (human) society ? Bonabeau et al
[1] explained it in the tradition of statistical physics, i.e. as
random. They claimed a phase transition between egalitarian society at
low population densities and hierarchical society at high population
densities. Sousa and Stauffer [2] found this transition to be an
artifact of the non-stationarity caused by their assumptions; and when the
rule was changed to give stationarity, the sharp transition
vanished. Now another simple rule change is shown to recover a phase
transition, though at a different population density.

\begin{figure}[hbt]
\begin{center}
\includegraphics[angle=-90,scale=0.50]{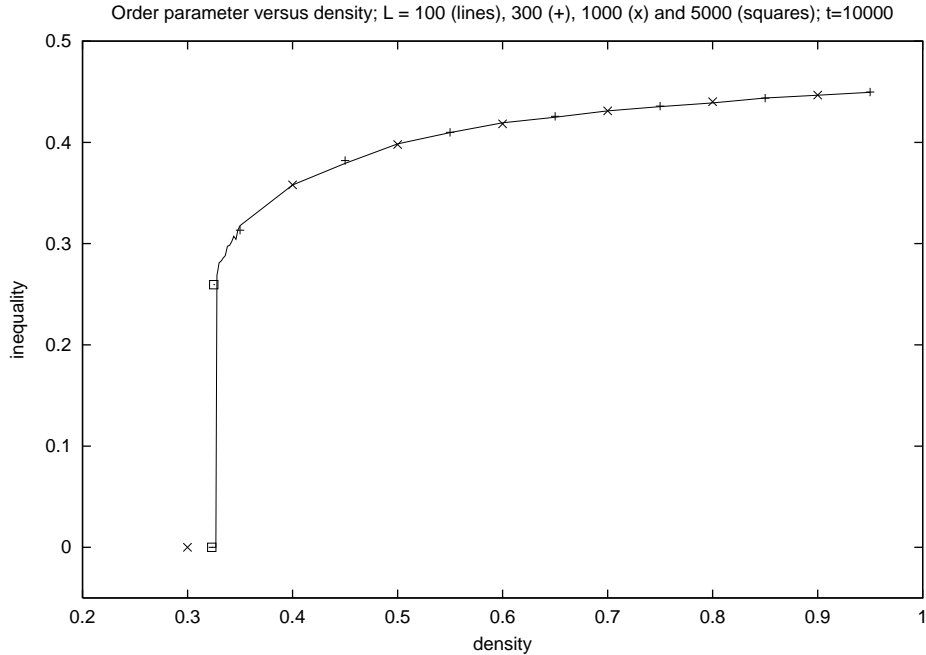}
\end{center}
\caption{
Variation of ``order parameter'' $\sigma$ versus population density
$p$, for $10^4$ iterations in $L \times L$ lattices, $L = 100$, 300, 1000
and 5000.
}
\end{figure}

In the model on Bonabeau et al [1] people initially are all equal and
are distributed randomly on a square lattice such that no two people
occupy the same place, filling a fraction $p$ of the lattice sites. 
They then diffuse randomly, and whenever one person $i$ wants to move
onto the site of person $k$ a fight between these two people
occurs. If $i$ wins, then $i$ and $k$ exchange their sites on the
lattice; if instead $k$ wins, none of the two moves. (The population density
is thus kept constant during the whole evolution, and is used as an
extrinsic parameter determining the final behaviour of the society: 
egalitarian or hierarchical.) The probability $q$ for $i$ to win is given by 
$$q = 1/(1 + \exp(\eta [h(k)-h(i)]) \eqno (1)$$
where $\eta$ is a free parameter and $h(j)$ counts the weighted number of
victories, minus the weighted number of losses, of person $j$. This
weight is initially unity and then decreases by ten percent at each
iteration (= one attempt to move on average all people through random 
sequential updating). Thus the memory of past fights fades away according
to an exponential rate, i.e. a finite-time memory. For
low population density the memory is lost between the rare collisions;
for high density the memories of the now frequent fights add
up. Simulations [2] gave a smooth increase of the amount $\sigma$ of
inequality with increasing density $p$. This amount of inequality is measured by
$$\sigma = (<q^2> - <q>^2)^{1/2} \eqno (2)  $$
and initially is zero if all $q$ are 1/2 (since all $h = 0$). 

To get a sharp transition, instead of a smooth variation, in
$\sigma(p)$, we now replace $\eta$ in Eq.(1) by $\sigma$, i.e. we
introduce a feedback of the order parameter $\sigma$ to the rules of
the fights [3]:
 $$q = 1/(1 + \exp(\sigma[h(k)-h(i)]) \quad .\eqno (3)$$
In reality this means that losing or winning has more
influence on the future in an hierarchical than in an egalitarian
society. (For the first ten iterations, the factor $\sigma$ is omitted
since then no inequalities had yet the chance to build up.) Fig.1
shows our results for lattices of sizes $100 \times 100 \dots 5000
\times 5000$, with a rather clear transition at a critical point
$p_c \simeq 0.32$. Fig.2 shows better data (averages over typically 64
lattices) close to the transition; it seems the order parameter jumps
from zero to about 1/4 when $p$ increases  above $p_c$.
 
We thank Suzana and Paulo Murilo de Oliveira for discussion, G. Weisbuch
for help during a computer breakdown, and the J\"ulich supercomputer center
for time on their Cray-T3E.

\begin{figure}[hbt]
\begin{center}
\includegraphics[angle=-90,scale=0.50]{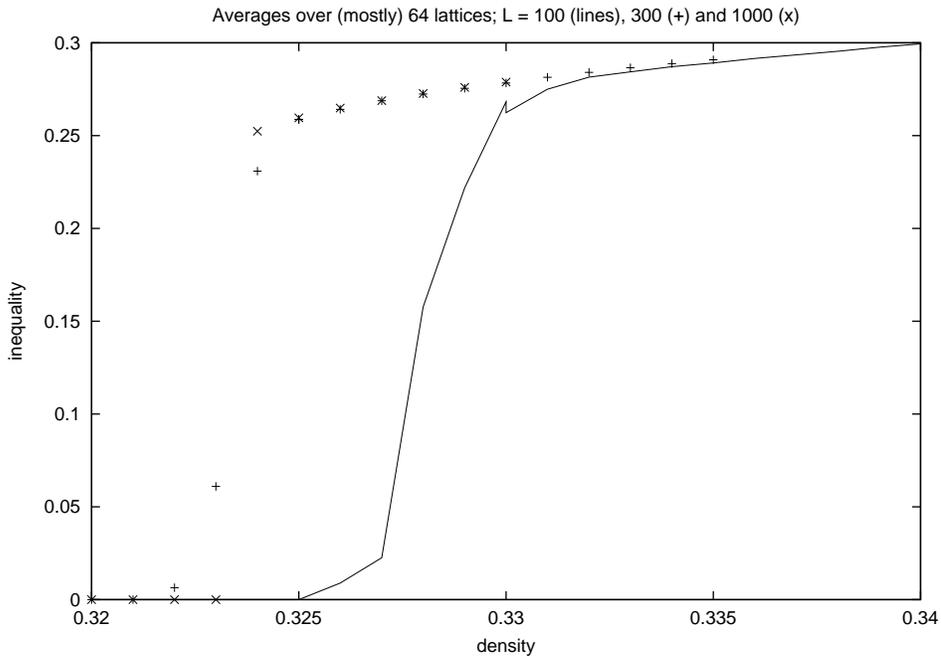}
\end{center}
\caption{
As Fig.1, but restricted to $p \simeq p_c$ and with better statistics.
Note the systematic downward shift of $p_c(L)$ with increasing lattice
size $L$.
}
\end{figure}

\bigskip
\parindent 0pt
[1] E. Bonabeau, G. Theraulaz, J.L. Deneubourg, Physica A 217, 373 (1995)
 
[2] A.O. Sousa, D. Stauffer, Int. J. Mod. Phys. C 11, 1063 (2000)

[3] S. Bornholdt, Int. J. Mod. Phys. C 12, 667 (2001)

\end{document}